\documentclass[superscriptaddress,showpacs,twocolumn,aps,pre,10pt]{revtex4-1}
\usepackage{natbib}

\usepackage{graphicx,dcolumn,bm,amsmath,color,amssymb,relsize}
\usepackage{units}
\usepackage{dsfont}
\usepackage{enumitem}
\usepackage[T1]{fontenc}

\newcommand{\eq}[1]{Eq.~\eqref{#1}}
\newcommand{\fig}[1]{Fig.~\ref{#1}}

\newcommand{\br}{ {\bf r} }

\newcommand{\bv}{ {\bf v} }
\newcommand{\bk}{ {\bf k} }
\newcommand{\bU}{ {\bf u} }

\newcommand{\gc}{\ensuremath{\gamma}}

\newcommand{\abs}[1]{\ensuremath{\left| #1 \right|}}
\newcommand{\F}{\ensuremath{\mathbf{F}}}
\newcommand{\T}{\ensuremath{\mathbf{T}}}

\newcommand{\avg}[1]{\ensuremath{\left\langle #1\right\rangle}}

\newcommand{\R}{\ensuremath{\mathbf{r}}}
\newcommand{\RR}{\ensuremath{\mathbf{R}}}

\newcommand{\kbt}{k_{\rm B}T}

\begin{document}
\title{Emerging activity in bilayered dispersions with wake-mediated interactions}

\author{J\"org Bartnick}
\email{bartnick@thphy.uni-duesseldorf.de}
\affiliation{Institut f\"ur Theoretische Physik II: Weiche Materie,
Heinrich-Heine-Universit\"at D\"{u}sseldorf, D-40225 D\"{u}sseldorf, Germany}

\author{Andreas Kaiser}
\affiliation{Materials Science Division, Argonne National Laboratory, Argonne, Illinois 60439, USA}
\author{Hartmut L\"owen}
\affiliation{Institut f\"ur Theoretische Physik II: Weiche Materie, Heinrich-Heine-Universit\"at D\"{u}sseldorf, D-40225
D\"{u}sseldorf, Germany}

\author{Alexei Ivlev}
\affiliation{Max-Planck-Institut f\"ur Extraterrestrische Physik, D-85741 Garching, Germany}

\date{\today}

\pacs{82.70.Dd, 52.27.Lw, 61.20.Ja, 05.40.Jc}

\begin{abstract}
In a bilayered system of particles with wake-mediated interactions, the action-reaction symmetry for
the effective forces between particles of different layers is broken. Under quite general conditions we show that, if the
interaction nonreciprocity exceeds a certain threshold, this creates an active dispersion of self-propelled clusters of
Brownian particles. The emerging activity promotes unusual melting scenarios and an enormous diffusivity in the dense fluid.
Our results are obtained by computer simulation and analytical theory, and can be verified in experiments with colloidal
dispersions and complex plasmas.
\end{abstract}

\maketitle

\section{Introduction}
\label{sec:intro}

Effective forces between mesoscopic particles often become nonreciprocal when the interactions are mediated by a
nonequilibrium environment. Such situations can be realized in various soft matter systems -- most notably in colloidal
dispersions~\cite{Hayashi2006,Buenzli2008,Dholakia2010,Shanblatt2011,Sabass2010,soto2014} and complex plasmas~\cite{Lowen2012,Bonitz2010,Morfill2009,Chaudhuri2011}, where microparticles are embedded, respectively, in a liquid solvent or a dilute weakly
ionized gas. In particular, the action-reaction symmetry in these systems is broken when the surrounding fluid moves with
respect to the particles~\cite{Melzer1996,Ivlev2015,Khair2007,Mejia-Monasterio2011,Dzubiella2003}, or when the interaction of molecules with the particle surface is out of equilibrium~\cite{Hayashi2006,Buenzli2008,Sabass2010,soto2014}.

Studies of nonreciprocal interactions have gained increased interest in recent time. When the dynamics of individual
particles is undamped (Newtonian) or weakly damped (when the relevant dynamical timescales are much shorter
than the damping time)~\cite{Ivlev2015}, which is typical for complex plasmas, one can observe a remarkable state of {\it
detailed dynamic equilibrium} with different species having different temperatures. For Brownian dynamics, it has recently
been shown that mixtures of diffusiophoretic colloids experience effective nonreciprocal forces which stimulate the
formation of stable aggregates (so-called active molecules)~\cite{soto2014} and trigger collective oscillatory
motion~\cite{soto2015}.

 In this paper, we consider a broad and generic class of nonreciprocal interparticle forces, the so-called
wake-mediated interactions. Particles embedded in a flowing medium generate wakes, which contribute to the interactions with
neighbors in a nonreciprocal way \cite{Melzer1996,Chaudhuri2011,Morfill2009}. Similarly, particles emitting chemicals in a
certain direction generate asymmetric concentration fields -- artificial ``chemical wakes'', also exerting non-reciprocal
forces on the neighbors \cite{soto2015,Bechinger}. In both cases, the action-reaction symmetry is only restored for
identical particles, forming a perfect monolayer perpendicular to the direction of wakes. Therefore, here we consider a
quasi two-dimensional system of Brownian particles~\cite{Leunissen2005} which are kept into stable bilayers by external
fields such as electric, magnetic, gravitational or optical fields. Under quite general conditions posed on the mutual
reciprocal and nonreciprocal forces, we observe a continuous transition from inactive (stacked) pairs to active units,
indicating the emergence of active fluids. Different from ordinary active particle
systems~\cite{Romanczuk2012,GompperRev,Bechinger}, these active units can break and become passive again. Using analytical
theory and simulation including hydrodynamic interactions between the particles, we explore the full density regime up to
freezing and find an unusual melting upon densification, along with a reentrant freezing and an enormous diffusivity in the
concentrated fluid.

The paper is organized as follows:  In Sec.~\ref{sec:model} we specify our model, perform a stability analysis for small clusters in Sec.~\ref{sec:stability} and describe our simulation in Sec.~\ref{sec:simu}. Results are discussed in Sec.~\ref{sec:Results} and summarized in Sec.~\ref{sec:conc}.

\section{Model}
\label{sec:model}

The motion of a particle $i$ at position $\R_i$ in the plane is governed by the fully damped
Langevin equation~\cite{Doi1986}
\begin{equation}\dot{\R}_i = \sum_{j} {\bf L}_{ij} ~\left(\F^{}_{j} + \boldsymbol{\xi}_j\right) + \frac{1}{2} \kbt \sum_j \frac{\partial {\bf L}_{ij}}{\partial \R_j},
\label{eom}
\end{equation}
where ${\bf L}_{ij}$ is the mobility matrix and $\boldsymbol{\xi}_i$ is a random force. The total force $\F^{}_{i}$ on
particle $i$ is given by $\F^{}_{i}= \sum_j\F^{}_{ji}$, where $\F^{}_{ji}$ is the pair-interaction force exerted by a
particle $j$ on the particle $i$. The random force $\boldsymbol{\xi}_i$ is Gaussian distributed with zero mean, $\langle
\boldsymbol{\xi}_i(t)\rangle = 0$, and variance $\langle \boldsymbol{\xi}_i(t) \boldsymbol{\xi}_j( t')\rangle = 2~
{\bf L}^{-1}_{ij}~ \kbt  \delta(t- t')$, where $T$ is the thermostat temperature, $k_{\mathrm B}$ the Boltzmann constant,
$\delta(t)$ the Dirac delta function, and ${\bf L}^{-1}$ the inverse of ${\bf L}$. In this paper, we include hydrodynamic
interactions in the zero-temperature limit, and neglect them at finite temperatures. The latter approach is
justified when the suspension is highly dilute, but still strongly interacting. Then, each mobility matrix reduces to
${\bf L}_{ij} = \gamma_i^{-1}\delta_{ij}{\bf I} $, with the unit matrix ${\bf I}$ and a friction coefficient $\gamma_i$.
In the zero-temperature limit, we consider the mobility matrix to be approximated by the Oseen tensor~\cite{Doi1986}
\begin{equation}
{{\bf L}_O}(\R) =  \frac{3 R_H}{4  \gamma_i r} ( {\bf I} + \hat{\R} \hat{\R}),
\label{eq:oseen}
\end{equation}
where $R_H$ is the hydrodynamic radius, $r = \abs{\R}$ and $\hat{\R} = \R / r$. Thus, we have ${\bf L}_{ij} \approx
{\bf L}_O(\R_i- \R_j)$ for $i \neq j$ and ${\bf L}_{ii} = {\bf I} / \gamma_i$.

We consider a typical situation when interactions between particles are isotropic in the plane. In this
case the mutual forces between particles $i$ and $j$ are radial, i.e., ${\bf F}_{ij}=F_{ij}{\bf n}_{ij}$ with ${\bf n}_{ij}$
being the unit vector from $i$ to $j$, and $F_{ij}$ only depends on the absolute distance $r_{ij}=|{\bf r}_i-{\bf r}_j|$.
Furthermore, we introduce species A and B and attribute particles to the same species if their pair interactions are
reciprocal, i.e., $F_{\mathrm{AA}}(r) =F_{\mathrm{BB}}(r)= -d\varphi_{\rm r}(r)/dr$. A generic form for the forces between
different species,
\begin{equation}\label{F_AB}
  F_{\mathrm{AB,BA}}(r) = -d\varphi_{\rm r}(r)/dr\pm d\varphi_{\rm n}(r)/dr,
\end{equation}
is a superposition of the reciprocal (r) and nonreciprocal (n) components, determined by the respective potentials
$\varphi_{\rm r,n}$. The latter are related to the potential $\varphi_{ij}$ generated by the particle $i$ at the location of
the particle $j$ via $\varphi_{\rm r,n}=\frac12(\varphi_{ji}\pm\varphi_{ij})$. Thus, the pair interactions are reciprocal if
$\varphi_{ij}=\varphi_{ji}$, and are nonreciprocal otherwise.

An important class of a constant nonreciprocity is realized when $\varphi_{\rm r}(r)$ and $\varphi_{\rm n}(r)$ are similar
functions, i.e., when the nonreciprocity $\varphi_{\rm n}(r)/\varphi_{\rm r}(r)\equiv\Delta ={\rm const}$.
For the undamped (Newtonian) or weakly damped dynamics with $\Delta={\rm const}$, the equations of motion
can be equivalently transformed into a reciprocal form by a simultaneous proper renormalization of the interaction forces
and masses, i.e., such dynamics can in fact always be described by a (pseudo) Hamiltonian~\cite{Ivlev2015}. It is noteworthy
that for the Brownian dynamics with nonreciprocal interactions one can employ a similar approach: By renormalizing the
interactions with Eq.~(3) of Ref.~\cite{Ivlev2015}, and introducing the renormalized damping coefficients
$\tilde\gamma_{\mathrm{A,B}} =\gamma_{\mathrm{A,B}}/(1\mp\Delta)$, we readily transform Eq.~(\ref{eom}) to the form where
the interactions are reciprocal, while the solvent temperatures for different species A and B are
different and equal to $\widetilde{T}_{\mathrm{A,B}} = T /( 1\mp\Delta)$. Interestingly, such a
``hetero-Brownian'' model has been recently introduced in a different context, to describe DNA
dynamics~\cite{Awazu2014a,Ganai2014}, and was also proposed for colloidal pairs under external forcing~\cite{Berut2014,GrosbergPRE2015,WeberPRL16}.

\begin{figure}
\includegraphics[width=\columnwidth]{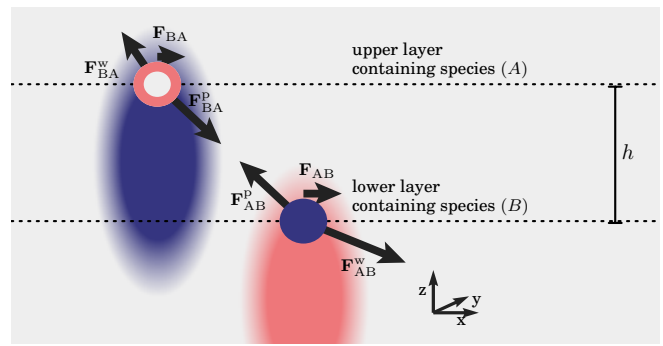}
\caption{Schematic sketch of nonreciprocal wake-mediated interactions. The particle species A and B are
confined in the upper and lower layers, respectively. While the direct interparticle forces are reciprocal,
$\F^{\rm p}_{\mathrm{BA}}+\F^{\rm p}_{\mathrm{AB}}=\mathbf{0}$, the particle-wake forces are nonreciprocal, $\F^{\rm w}_{\mathrm{BA}}+\F^{\rm w}_{\mathrm{AB}}\neq \mathbf{0}$
such that for the total forces $\F_{\mathrm{AB}} \neq - \F_{\mathrm{BA}}$.}
\label{figintro}
\end{figure}

In our model, we adopt a binary mixture of point-like particles whose direct (electrostatic) interactions
are characterized by the charges $Q_{\mathrm{A}}$ and $Q_{\mathrm{B}}$. The particles are confined in a horizontal
$xy$-plane in two layers with a height difference $h$, as sketched in \fig{figintro}. The point-like approximation is
justified as long as the distance between the particles is much larger than their diameter. For non-reciprocal particle-wake
forces we have two representative realizations in mind, both leading to effective Yukawa interactions: The first one can be
obtained with catalytically active Janus colloids~\cite{Howse_2007,ZhengBtH2013}, where each particle emits a chemical from
the lower segment of the surface, thus generating ``chemical'' wakes (the vertical orientation of such particles can be
stabilized in an external magnetic field~\cite{Baraban2012}). The emitted chemicals decompose with a constant
rate~\cite{GrimaPRL05}, such that the effective interaction between the chemical wake and a neighboring particle is of
Yukawa form. The second realization is an externally imposed micro-ion flow, parallel to the vertical $z$-axis, which
induces an ``electrostatic'' wake below each particle, while the fluid remains at rest. In both cases, the wake is mimicked
by a point-like effective ``charge'' $q_i$ at the distance $\delta$ below each particle.

The total interaction force is a combination of the direct particle interaction and the particle-wake interaction. The force
depends on several free parameters whose combination determines emergence of different self-organization phenomena. In this
Letter we chose a certain set of parameters, demonstrating variety of the emerging activity, and identify a general
necessary condition for the activity onset [see Eq.~(\ref{ConditionActiveDoublet}) below]. Let us introduce the
three-dimensional particle coordinates $\RR_i$ and the corresponding coordinates $\R_i$ in the horizontal plane. Then the
force ${\bf F}_{ij}= -\partial\varphi_{ij} /\partial{\R}_j$ exerted in the horizontal plane by the particle $i$ on the
particle $j$ is determined by the potential $\varphi_{ij}=Q_iQ_jY(R_{ij})+q_iQ_jY(R_{ij}^{\rm w})$, where
$Y(R)=R^{-1}e^{-R/\lambda}$ is the (unity charge) Yukawa potential which depends on the distance $R_{ij}=|\RR_{i}-\RR_j|$
between the particles as well as on the distance $R_{ij}^{\rm w} = |\RR_{i}-\RR_j - \delta \mathbf{n}_z|$ between the
particle $j$ and the wake center of the particle $i$; here, we assume that a Yukawa potential for both forces has the same
effective screening length $\lambda$, and that $q_i\propto-Q_i$~\cite{plasmafootnote,Couedel2010,Ivlev2000}. For particles
in the same layer, A or B, we have $R_{ij}^{\rm w}=R_{ji}^{\rm w}$; therefore, $\varphi_{\rm n}=0$ (since
$q_iQ_j=q_jQ_i$), and hence the forces are reciprocal. For the AB interactions the symmetry is broken, $R_{ij}^{\rm w}\ne
R_{ji}^{\rm w}$, and the forces are nonreciprocal. For simplicity, particles A and B have charges of the same magnitude and
opposite signs, $Q_{\mathrm{A}}=-Q_{\mathrm{B}} \equiv Q$, the same friction coefficients,
$\gamma_{\mathrm{A}}=\gamma_{\mathrm{B}}\equiv \gamma$, and the height difference is $h=\lambda$. A natural measure of
nonreciprocity in this case is the relative wake charge, $\tilde q = -q_i / Q_i>0$.

\section{Stability analysis for small clusters}
\label{sec:stability}

In order to illustrate a tendency of particles with nonreciprocal interactions to self-organize
themselves with increasing $\tilde q$, and to identify the characteristic building blocks of this complex process, let us
consider the formation of small clusters in the absence of hydrodynamic interactions. Then, the equilibrium configurations
for a cluster of $N$ particles are determined from the force balance in the {\it horizontal} plane, 
\begin{equation}
\sum_j^N\F_{ji}(r_{ij})= \F,
\label{eq:ForceBalance} 
\end{equation} 
where the net force $\F$ is a constant horizontal vector for $\forall i \in [1,N]$. We apply the standard stability
analysis of the derived configurations in the zero-temperature limit. This corresponds to the eigenvalue problem
$\det\left(\partial \F_{ij}/\partial \R_{j}|_{\mathrm{eq}} - \gc \omega {\bf I}\right) = 0$, where $\ldots|_{\mathrm{eq}}$
denotes the ($2N\times2N$) dynamical matrix calculated for the equilibrium configurations.

\subsection{Doublets}

A pair of particles of different species form an equilibrium {\it doublet} with the horizontal separation $r_D$ when
$F_{\mathrm{AB}}(r_D)=-F_{\mathrm{BA}}(r_D)\equiv F$; the doublet is stable if $\left.d\left[F_{\mathrm{AB}}(r) + F_{\mathrm{BA}}(r)
\right]/dr\right|_{r=r_D} <0$.
From Eq.~(\ref{F_AB}) we conclude that the stability condition is only fulfilled when the
reciprocal component of the force is equal to zero, $\left.d\varphi_{\rm r}(r)/dr\right|_{r=r_D} = 0$.

\begin{figure}
\includegraphics[width=\columnwidth]{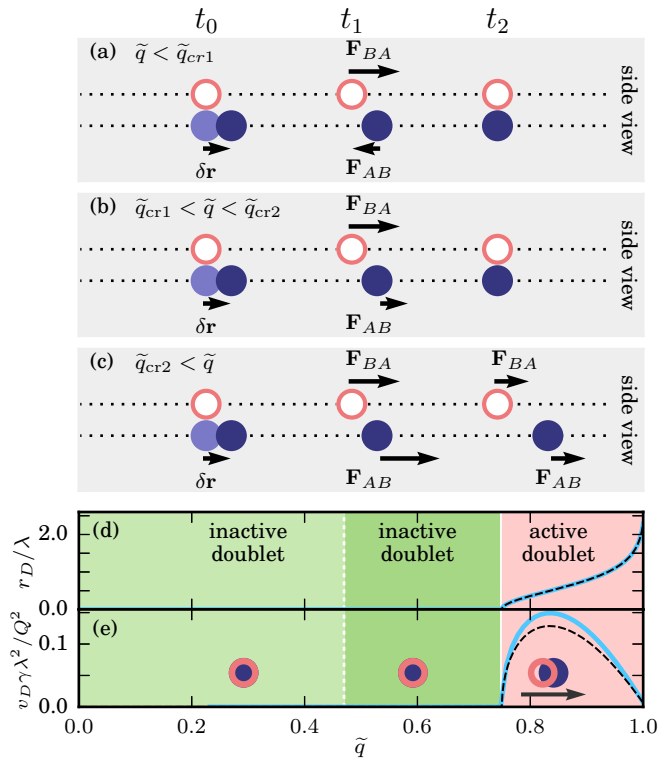}
\caption{Stable configurations of two-particle clusters, depending on the relative wake charge $\tilde q$.
(a-c) Sketches illustrate three distinct regimes (side view): For $\tilde q< \tilde q_{\rm cr1}$, particles form a stable
vertical pair, {\it an inactive doublet}, since the restoring forces ${\bf F}_{\mathrm{AB}}$ and ${\bf F}_{\mathrm{BA}}$ exerted by a small
perturbation pull the particles back; for $\tilde q_{\rm cr1}<\tilde q< \tilde q_{\rm cr2}$, the vertical
pair remains stable only in the zero-temperature limit assumed here, since ${\bf F}_{\mathrm{AB}}$ and ${\bf F}_{\mathrm{BA}}$ are pointed in the same
direction; for $\tilde q> \tilde q_{\rm cr2}$, the particles form {\it an active doublet} with a finite horizontal
separation, moving along the force ${\bf F}_{\mathrm{AB}}={\bf F}_{\mathrm{BA}}$. (d,e) Equilibrium horizontal separation of the doublet,
$r_D$ (normalized by $\lambda$) and the corresponding doublet velocity $v_D$ (normalized by $Q^2/\lambda^2\gamma$), the
shading indicates the stability regimes illustrated in (a-c).
The results are for the wake length $\delta=0.2\lambda$, $h = \lambda$ and $R_H =0$ (black dashed line) as well as $R_H=0.2\lambda$ (blue solid line).}
\label{figtheory}
\end{figure}

For a vertical pair $r_D = 0$ -- we call it {\it an inactive doublet}
-- two regimes can be distinguished: (i) When $\left. dF_{ij}(r) /dr\right|_{r=0} <0$ for both particles, they return to the
equilibrium after a small perturbation. Below we demonstrate that this case, sketched in \fig{figtheory}(a), is observed for
a ``weak'' nonreciprocity, when the relative wake charge is smaller than a certain critical value, $\tilde q< \tilde q_{\rm
cr1}$ (i.e., this always occurs for reciprocal interactions). (ii) When $\left. dF_{ij}(r) /dr\right|_{r=0}>0$ for one of
the particles, the restoring forces are pointed in the same direction, as shown in \fig{figtheory}(b). The equilibrium in
this case, corresponding to $\tilde q_{\rm cr1}<\tilde q< \tilde q_{\rm cr2}$, would only be restored in the
zero-temperature limit; in the presence of an infinitesimal thermal noise the doublet should break apart.

Under the general condition
\begin{equation}
\left.d\varphi_{\rm r}(r)/dr\right|_{r=r_D} = 0\,\,\, {\rm and}\,\, \left.d\varphi_{\rm n}(r)/dr\right|_{r=r_D} \neq 0,
\label{ConditionActiveDoublet}
\end{equation}
satisfied for $r_D > 0$, a pair emerges which is self-propelled in the direction ${\bf n}_{\mathrm{AB}}$ with the velocity
$v_D=- \gamma ^{-1}\left.d\varphi_{\rm n}(r)/dr\right|_{r=r_D}$. Such clusters will be referred to as {\it active doublets}
and occur when $\tilde q> \tilde q_{\rm cr2}$. Note that for a constant nonreciprocity, $\Delta={\rm const}$, stable
doublets are always at rest, since $\varphi_{\rm n}(r)=\Delta\varphi_{\rm r}(r)$ and therefore the nonreciprocal force is
equal to zero at $r=r_D$~\cite{fn}. Equation~(\ref{ConditionActiveDoublet}) represents the necessary condition for the emerging activity, which
can be satisfied for various combinations of the interaction parameters (e.g., when $Q_{\mathrm{A}}$ and $Q_{\mathrm{B}}$
have the same sign, but the direct and the particle-wake interactions are characterized by different screening lengths).

Figures~\ref{figtheory}(d) and (e) illustrate the results of the stability analysis in the horizontal plane, performed in
the zero-temperature limit. In the present example, two particles of different species are stacked on top of each other
(i.e., they form an inactive doublet) when the relative wake charge is smaller than $\tilde q_{\rm cr2} \simeq 0.74$. For
larger $\tilde q$, the separation $r_D$ continuously increases and an active doublet moves along its symmetry axis, with the
velocity $v_D$ which varies non-monotonically with $\tilde q$. Thus, in dilute systems (with infinitesimal number density)
one can expect the formation of multiple individual doublets. Hydrodynamic interactions do not influence the pair separation
$r_D$, but cause a velocity increase. Interestingly, the leading term for the hydrodynamic far-field for an active doublet
is a force monopole, as opposed to a standard microswimmer where it is a force dipole~\cite{GompperRev}.

For a finite number density $\rho$ (number of particles per unit area) in systems without hydrodynamic interactions, we
analyze the stability of crystalline structures in the zero-temperature limit. The time-dependent coordinate of the
$i$th-particle is presented as a sum of its equilibrium lattice position and a displacement, $\R_i(t) = \R_{{\rm
eq},i}+\bU_i(t)$. The interaction force, \eq{F_AB}, is then expanded to the first order in $\bU_i$ and substituted in
\eq{eom}. Using $\bU_i \propto \exp (i \bk \cdot \R_{{\rm eq},i} + \omega t)$, the dispersion relations $\omega({\bf k})$
are derived as eigenvalues of the resulting dynamical matrix~\cite{Couedel2010}. We examine a vertically stacked hexagonal
lattice and an interdigitated hexagonal lattice, the stability requires Re~$\omega({\bf k}) < 0$ for all ${\bf k}$ from the
first Brillouin zone of the lattice.

Let us now study the effect of hydrodynamic interactions on a doublet, consisting of a particle of 
species A and one of species B in the dilute limit. 
The general requirement for stable clusters is the equality of the velocities,
\begin{equation}
\dot{\R}_{\mathrm{A}} = \dot{\R}_{\mathrm{B}}\, .
\label{eq:hydro_crit_doublet}
\end{equation}
The equation of motion of the A particle can be written as 
\begin{equation}
\dot{\R}_{\mathrm{A}} = \frac{1}{\gamma} \F_{\mathrm{A}}(\R_{\mathrm{A}} - \R_{\mathrm{B}}) + \left[ \frac{3 R_H}{4 \gamma \tilde{r}} \left( 1 + \frac{r^2}{\tilde{r}^2}\right)\right] \F_{\mathrm{B}}(\R_{\mathrm{B}} - \R_{\mathrm{A}}) \, ,
\label{eq:hydro_eom}
\end{equation}
with $\tilde{r} = \sqrt{r^2 + h^2}$;
the respective equation for the  B particle, is obtained by the $\mathrm{A} \leftrightarrow \mathrm{B}$ permutation.
Equation~\eqref{eq:hydro_crit_doublet} is only fulfilled if $F_{\mathrm{AB}}(r_D)=-F_{\mathrm{BA}}(r_D)$, as in the case without hydrodynamic interactions.
Thus, Eqs.~(\ref{ConditionActiveDoublet}) remain valid also for a finite hydrodynamic radius.
Figure~\ref{figtheory} shows the result for a doublet with and without hydrodynamic interactions. 
The doublet distance $r_D$ remains unchanged, while the doublet velocity is slightly increased.

\subsection{Triplets}

\begin{figure}
\includegraphics[width=.5\textwidth]{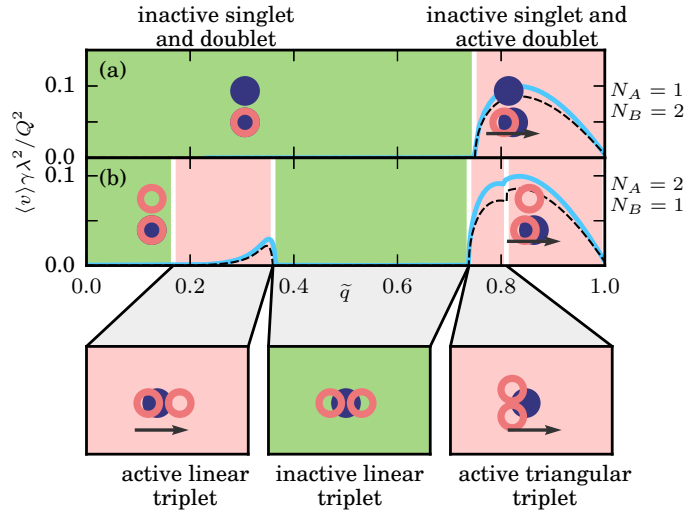}
\caption{ Stable configurations and velocities of three-particle clusters in the zero-temperature limit. The two possible combinations of the species A and B are shown. The figure legend is the same as in Fig.~\ref{figtheory}.}
\label{figtheoryT}
\end{figure}

For three particles, there is a variety of possible {\it triplet} configurations. 
To start with, let us consider a cluster composed of one particle B and two particles A with negligible hydrodynamic interactions.
We work in the frame of reference of the first (B) particle, i.e., the coordinates of the second and third (A) particles are ${\bf r}_{2,3}$. 
In this case, the general equilibrium condition, Eq.~(\ref{eq:ForceBalance}),
can be identically transformed to the following two equations for the particle coordinates (plus one equation for the net
force $\F$):
\begin{subequations}
\begin{align}
 \tilde F(r_{2}){\bf n}_2 + \tilde F (r_{3}){\bf n}_3={\bf 0}, \label{maineqa}  \\
  2F_{\mathrm{AA}} (r_{23}){\bf n}_{32} -F_{\mathrm{BA}}(r_{3}){\bf n}_3 + F_{\mathrm{BA}} (r_{2}){\bf n}_2={\bf 0},  \label{maineqb}
\end{align}
%\label{maineq}
\end{subequations}
\noindent with $\tilde F(r)\equiv F_{\mathrm{BA}}(r) + 2F_{\mathrm{AB}}(r)$. 
In the reverse case, where clusters are composed of one A and two B particles, the labels are simply to be swapped.

Using \eq{maineqa}, one can distinguish two principal cases: (i) $\tilde
F(r_{2,3}) \neq 0$, then solutions exist only for $\R_2 \parallel \R_3$; (ii) $\tilde F(r_{2}) = \tilde F(r_{3}) = 0$, then
solutions are possible for noncollinear $\R_2$ and $\R_3$. 

\begin{enumerate}[label=(\roman*)]
\item If $\tilde F(r)$ is a monotonic function, the only solution is $\R_2 =-\R_3$; from \eq{maineqb} we obtain
    $F_{\mathrm{AA}}(2r) = F_{\mathrm{BA}}(r)$, which yields $r_2 = r_3 \equiv r$. Due to symmetry, $F=0$ and hence we call such
    configurations {\it inactive linear triplets}. However, if $\tilde F(r)$ is a non-monotonic function, also solutions
    with $r_2 \neq r_3$ are possible --  in this case Eqs.~(\ref{maineqa}) and~(\ref{maineqb}) are reduced to $\tilde F(r_2)=\tilde F(r_3)$
    and $2 F_{\mathrm{AA}}(r_2+r_3) = F_{\mathrm{BA}}(r_2)+F_{\mathrm{BA}}(r_3)$. Such asymmetric clusters usually imply a non-vanishing net force,
    $F\neq0$, which generates a directed propulsion. We call these configurations {\it active linear triplets}.
\item If $F_{\mathrm{BA}}(r)$ is monotonic, solutions for ${\bf r}_{2,3}$ are limited to triangles with $r_2 = r_3 \equiv r$ and
    the apex angle $\theta$, obtained from $F(r)=0$ and $F_{\mathrm{AA}}(2 r \sin \frac12\theta)/\sin \frac12\theta =F_{\mathrm{BA}}(r)$.
    Such configurations are called {\it active triangular triplets}. Finally, if $F_{\mathrm{BA}}(r)$ is a non-monotonic
    function, triangular triplets with $r_2\ne r_3$ are possible. 
\end{enumerate}
As for the doublets, we apply the standard stability analysis of the derived configurations in the zero-temperature limit.

%%%%%%%%%%%%%%%%%%%
\begin{figure}[t]
\centering
\includegraphics[width=.5\textwidth]{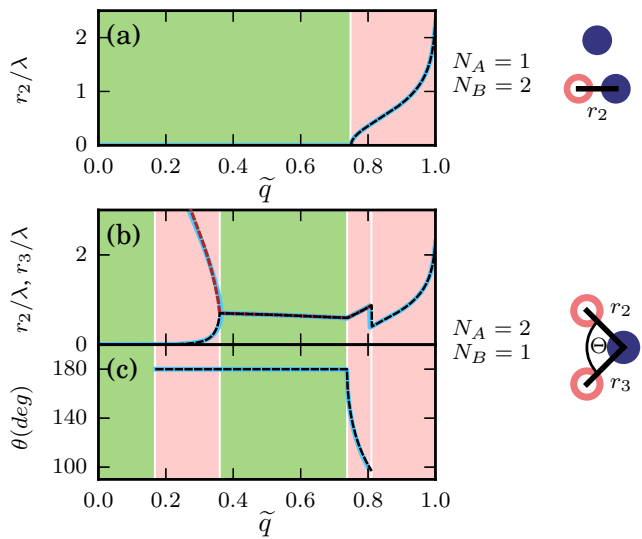}
\caption{Equilibrium horizontal separation in the doublet and triplet configurations shown in Fig.~\ref{figtheoryT}.  }
\label{SFig2}
\end{figure}
%%%%%%%%%%%%%%%%%%%%%%

If the number of particles B is twice as high as the number of A particles, the dependence on $\tilde{q}$ remains the same 
as in Fig.~\ref{figtheory}. The
``excess'' particle B simply remains an {\it inactive singlet} [see \fig{figtheoryT}(a)]. On the contrary, in the 
situation with two particles A for one particle B various active and inactive structures emerge, as presented in
\fig{figtheoryT}(b): An inactive doublet and an inactive singlet are formed when $\tilde{q} < 0.17$, while for $\tilde{q} \in (0.17, 0.36)$
they merge into an \emph{active linear triplet}, where the position of particle B is slightly shifted from the center (which
determines the propagation direction along the symmetry axis). The linear triplet becomes inactive at $\tilde{q} \in (0.36,
0.74)$. The further increase of the wake charge, $\tilde{q} \in (0.74, 0.81)$, causes particle B to shift perpendicular to the
symmetry axis, leading to an \emph{active triangular triplet}. For even larger values of $\tilde{q}$, the triplet breaks apart and an
active doublet and an inactive singlet emerge. In a similar manner, one can straightforwardly generalize the analysis for
larger clusters or investigate, e.g., the rotation activity.

Figure~\ref{SFig2}(a) shows the horizontal separation $r_1$ for the case of $(N_A = 1, N_B = 2)$, where a passive singlet and a doublet form.
In the reverse situation $(N_A = 2, N_B = 1)$, we characterize the emerging triplets by their individual bond distances $r_2,r_3$ [Fig.~\ref{SFig2}(b)] and the respective apex angle $\theta$ [Fig.~\ref{SFig2}(c)]. 
Activity is a result of symmetry breaking, therefore active units are found if $r_2 \neq r_3$ or $\theta < 180^{\circ}$.

If hydrodynamic interactions are taken into account, the equilibrium condition generalizes towards %$\dot{\R}_i = \dot{\R}$.
\begin{equation}
\dot{\R}_i = \dot{\R} \quad \forall \quad  i\in \left[ 1, N\right]~.
\label{eq:hydro_crit_triplet}
\end{equation}
Generally, the triplet coordinates derived above do not fulfill this equilibrium condition,
which is in contrast with the doublets (where the inclusion of hydrodynamic interactions only induces a rescaling of the velocity).
For three particles, we solve Eq.~(\ref{eom}) numerically and show the results in Figs.~\ref{figtheoryT} and \ref{SFig2}.
However, the resulting changes to the particle coordinates are minor, as shown in Fig.~\ref{SFig2}.
Similar to doublets, the inclusion of hydrodynamic interactions merely causes a slight increase of the velocity, see again Fig.~\ref{figtheoryT}.

\section{Computer simulations}
\label{sec:simu}
We solve the equation of motion, Eq.~(\ref{eom}), using a forward time-step algorithm in a Brownian dynamics simulation for three distinct cases:
when the hydrodynamic interactions are neglected, we consider (i) the zero-temperature limit and (ii) finite temperatures; 
 (iii) the effect of the hydrodynamic interactions is studied in the zero-temperature limit.
We use a 2D rectangular simulation box with periodic boundary conditions and the edge ratio $L_y/L_x =\sqrt{3}/ 2$. 
The particles are initialized on a distorted stacked hexagonal lattice with a fixed number density $\rho=N/(L_xL_y)$. 
In the case (i) and (ii), we use $N=2 \times 2500$ particles.
The respective edge lengths of the simulation domain are varying from  $(L_x, L_y) \simeq (240\lambda, 210\lambda)$ 
at low densities to $(L_x, L_y) \simeq (43 \lambda,38 \lambda)$ at $\rho\lambda^2 = 3$. 
In the case (iii), we use $N = 2 \times 576$ particles,
and the simulation domain is between $(L_x, L_y) \simeq (115\lambda, 100\lambda)$ 
at low densities and $(L_x, L_y) \simeq (28 \lambda, 24 \lambda)$ at high densities.
The modeling of the hydrodynamic interactions is done on the Oseen level with $R_H = 0.2\lambda$, i.e. up to volume fractions of about $15\%$.
For all cases, the wake length is $\delta = 0.2 \lambda$, the height between the layers is $h = \lambda$, 
the time $t$ is measured in units of $\tau = \gc \lambda^3/Q^2$ and the distance $r$ in units of $\lambda$. 
We set the time step to $\delta t=0.005\tau$ in the cases (i) and (ii) and to  $\delta t=0.0025\tau$ in case (iii), 
which ensures proper resolution of the particle dynamics. 
After initialization, the system is given time of $10^4\tau$ to relax into a steady state. 
Statistics is gathered for multiple simulations runs with independent initializations and the simulation time of $2500\tau$. 
By measuring the displacement of individual particles within the time step,
a particle velocity is calculated as $\bv_i (t)= [\br_i(t + \delta t) - \br_i(t)]/ \delta t$.

\section{Results}
\label{sec:Results}

\subsection{Zero-temperature limit}

\begin{figure}
\includegraphics[width=\columnwidth]{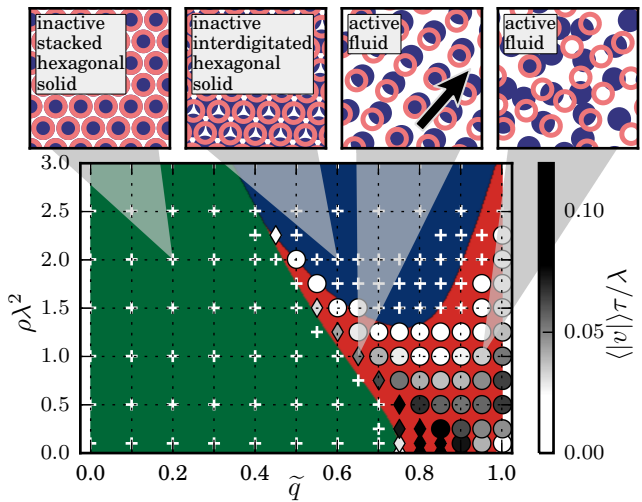}
\caption{State diagram in the zero-temperature limit, plotted in the plane of the number density $\rho$ and
relative wake charge
$\tilde q$. Color coding depicts results obtained from the stability analysis, symbols show numerical results. Inactive
systems ($+$) can be either {\it stacked hexagonal solid} (green background) or {\it interdigitated hexagonal solid} (blue
background). For {\it active fluid} regimes ($\bigcirc$, red background), the average particle velocities are
indicated by a gray scale. Diamonds ($\diamondsuit$) are used instead of circles if active doublets emerge whose decay
time $\tau_D$ exceeds a threshold of $10^3\tau$. The states are illustrated by typical snapshots, see also movies in the
Supplemental Material \cite{SuppMat}.}
\label{figv}
\end{figure}

\begin{figure}[tbhp]
\centering
\includegraphics[width=\columnwidth]{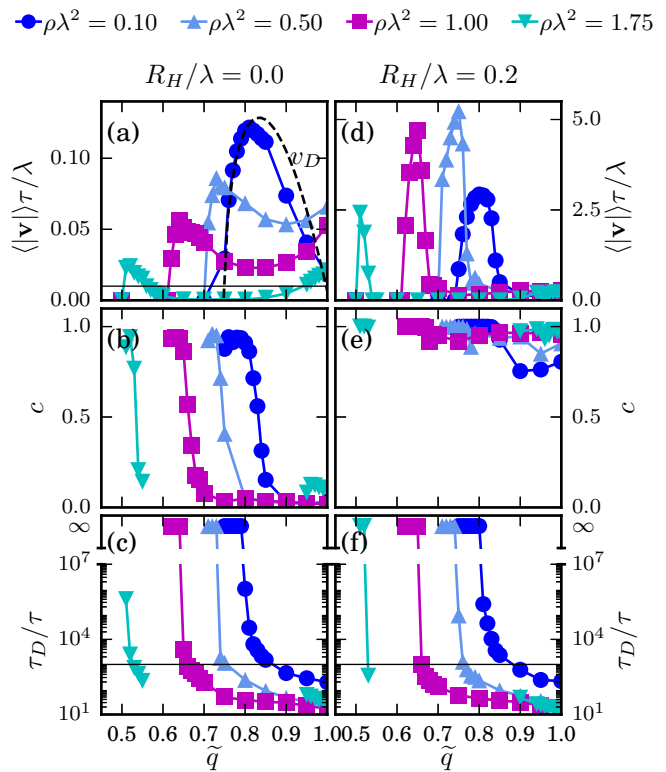}
\caption{Characteristics of active fluids: (a) and (d) average particle velocity $\left\langle |{\bf v}|
\right\rangle$, (b) and (e) alignment parameter $c$, and (c) and (f) decay time of doublets $\tau_D$, 
plotted versus the relative wake charge $\tilde q$ for several values of the number density $\rho$. The panels 
on the left show the results for simulations without hydrodynamic interactions, on the right hydrodynamic 
interactions are included. The dashed line in panel (a) shows the velocity of a
single active doublet $v_D$ in the dilute case ($\rho\lambda^2\ll1$), the horizontal lines in panels (a), (c) and (f) indicate
the threshold values of the velocity ($10^{-2}\lambda/\tau$) and decay time ($10^3\tau$).}
\label{figop}
\end{figure}

The above analytical results, see Sec.~\ref{sec:stability}, are complemented with a numerical analysis~\cite{Colberg2011}.
Figure~\ref{figv} presents the state diagram of the emerging activity, where we compare the theoretical results against
simulations in the zero-temperature limit. The state diagram is plotted in the plane spanned by number density $\rho$ and
relative wake charge $\tilde q$. We identify three distinct domains: Toward the reciprocal limit $\tilde q =0$, the
particles form a bilayered {\it stacked hexagonal crystal} (green); for larger $\tilde q$, at increased density the system
goes into an {\it interdigitated hexagonal solid} (blue); for even larger $\tilde q$, at low densities we find an {\it
active regime} where the crystal melts (red). In addition, we show the results obtained from the numerical simulations.
Here, we differentiate between {\it inactive solids} ($+$) and {\it active fluids} ($\bigcirc$). The mobile units of the
fluid are active doublets, that behave similar to (deformable) active Brownian
particles~\cite{MenzelEPL2012,PagonabarragaJML,StarkPRL2014}. The active regime in the simulations is defined for the
average particle velocity $\left\langle | {\bf v} | \right\rangle$ above a threshold of $10^{-2}\lambda/\tau$. One can see
that the emerging state diagram exhibits a reentrant behavior both with $\rho$ and $\tilde q$. Notably, for intermediate
$\tilde q$, there is an anomalous ``water-like'' melting upon an increase in $\rho$ followed by reentrant freezing.
Furthermore, to quantify the stability of doublets we define $N_D(t)$, the average number of particle pairs that remain
nearest neighbors over the time interval $t$. Generally, it is well described by an exponential decay, $N_D(t) \propto
e^{-t/\tau_D}$, with a doublet decay time $\tau_D$. Long-living active clusters are marked by a diamond in
\fig{figv}. The existence of a finite decay time $\tau_D$ reveals a qualitative difference of our system to a system of
permanently active particles \cite{GompperRev,Bechinger,MarchettiPRL2012,ReichhardtPRE2015}.

\begin{figure}[htb]
\begin{center}
\includegraphics[clip=,width=.485\textwidth]{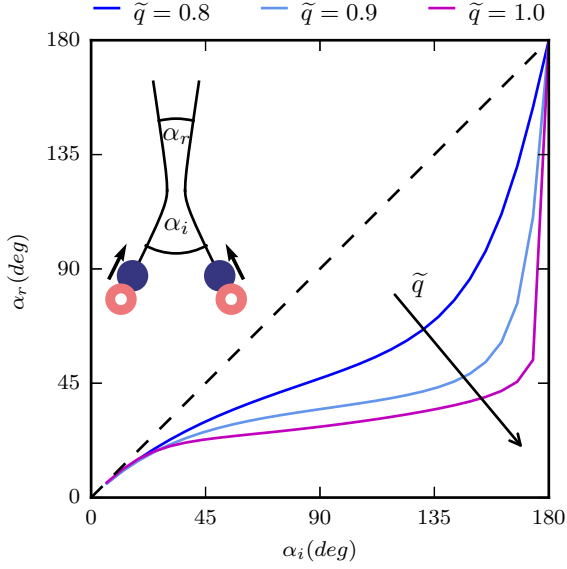}
\caption{\label{Collision} Angle of reflection $\alpha_r$ versus the incidence angle $\alpha_i$ for a collision of two active doublets, plotted for different values of the relative wake charge $\tilde{q}$.
The symmetric scattering, $\alpha_r = \alpha_i$, is indicated by the dotted line.}
\end{center}
\end{figure}

We now discuss the characteristics of the {\it active fluid} in more detail. Let us first introduce
the averaged velocity $\langle \bv \rangle = \langle[\br_i(t + \delta t) - \br_i(t)]/ \delta
t \rangle$, see \fig{figop}(a) and (d), as well as an alignment parameter $c=\abs{\avg{\bv}} / \avg{\abs{\bv}}$, see \fig{figop}(b) and (e): $c=1$ for  a perfect nematic order and $c=0$ in a totally disordered case. The stability
of doublets is quantified by $N_D(t)$, the averaged number of particle pairs that remain nearest neighbors over the time interval $t$. 
%By an exponential fit $N_D(t) \propto e^{-t/\tau_D}$, we can obtain 
The doublet decay time $\tau_D$ is shown in \fig{figop}(c) and (f). If no doublet splits during the simulation time of
$2500\tau$, then $\tau_D$ is set to infinity.

Figure~\ref{figop}(a) demonstrates that at low densities ($\rho\lambda^2=0.1$), the average velocity $\left\langle |{\bf v}|
\right\rangle$ is well reproduced by the velocity of a single active doublet, $v_D$, calculated analytically. Above the
threshold value of $\tilde q_{\rm cr2} = 0.74$, the distance $r_D$ increases, see Fig.~\ref{figtheory}(d). 
For this reason, the
average velocity first increases with $\tilde q$, but then it starts falling off due to decreasing interaction strength of a
doublet, see also Fig.~\ref{figtheory}(e). 
As the activity sets in, long-living doublets are formed throughout the system and
their mutual collisions lead to the velocity alignment, see Fig.~\ref{figv},
since the angle of reflection $\alpha_r$ after their mutual collision is always smaller the the incidence angle $\alpha_i$, as shown in
 Fig.~\ref{Collision}.
With increasing the number density $\rho$ the onset of activity shifts towards smaller $\tilde q$, whereby the average
velocity vs. wake charge becomes a non-monotonic function, leading to a reentrant effect for $\rho\lambda^2 > 1.25$, where
an {\it inactive interdigitated hexagonal solid} emerges, see Fig.~\ref{figv}.

The effects of hydrodynamic interactions are demonstrated in the right panel of \fig{figop}.
One can see that the effects become significant at higher densities. 
In comparison with the left panel, we observe a drastic velocity increase, while the alignment remains strong at any $\tilde q$.
These effects become more pronounced due to the long-range nature of hydrodynamic interactions, so that
more particles are involved in the collective motion.

%%%%%%%%%%%%%%%%%%

\subsection{Finite temperature study}

\begin{figure}
\includegraphics[width=\columnwidth]{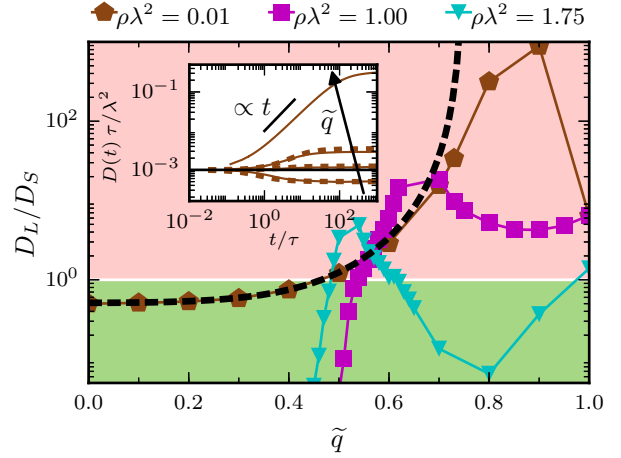}
\caption{
Ratio of the long-time to short-time diffusion coefficients, $D_L/D_S$, obtained from the
time dependent diffusion coefficient $D(t)$
for a finite temperature $T = 10^{-3}\, Q^2 / (k_{\rm B} \lambda)$. The shading indicates a transition between active fluids
($D_L/D_S>1$) and solids ($D_L/D_S<1$). The inset depicts the normalized $D(t)$ for $\rho \lambda^2 = 0.01$ and
$\tilde q \in \{0,0.5,0.6,0.8\}$, demonstrating the transition from subdiffusive to ballistic intermediate behavior with increasing
$\tilde q$. The dashed lines represent the analytical solutions of the Langevin equation for the diffusion of a doublet in the dilute limit, Eq.~(\ref{eq:diffratio}), the solid lines and symbols show numerical results.}
 \label{figMSD}
\end{figure}

Let us study the case of finite temperatures in the absence of hydrodynamic interactions. 
For a single stacked doublet, where $r_D = 0$, we compute the mean-squared displacement from Eq.~(\ref{eom}).
We take the Taylor expansion of the forces around the equilibrium positions.
Then, the radial force perturbations are  $\delta F_{\mathrm{BA}} (t) \approx C_{\mathrm{A}} ~[\delta r_{\mathrm{A}}(t) - \delta r_{\mathrm{B}}(t)]$ and $\delta F_{\mathrm{AB}} (t) \approx C_{\mathrm{B}} ~[\delta r_{\mathrm{B}}(t) - \delta r_{\mathrm{A}}(t)]$, where $C_{\mathrm{A}}$ and $C_{\mathrm{B}}$ are the prefactors of the linear-order terms in the expansion.
We define ${\bf A}$ as a matrix containing these prefactors, such that Eq.~(\ref{eom}) can be written as
\begin{equation}
\gamma \frac{\partial \mathbf{X}(t)}{\partial t} = {\bf A}(\tilde q) \mathbf{X}(t) + \mathbf{T}(t),
\end{equation} 
where $\mathbf{X}(t) = \left\{ \delta r_{\mathrm{A}}(t), ~\delta r_{\mathrm{B}}(t)\right\}$ is the vector of the particle positions
and $\T(t) = \left\{{\xi}_1(t),~{\xi}_2(t) \right\}$ the vector with the random (radial) forces acting on the particles. 
For simplicity, we set the friction coefficient $\gamma$ independent of the particle index.
Using variation of constants, this differential equation is solved by the integration over a matrix exponential:
\begin{equation}
 \mathbf{X}(t) = \frac{1}{\gamma} \int_0^t \mathrm{d}\tau ~ \exp\left[{\bf A}~ (t- \tau) / \gamma\right] ~ \T(\tau),
 \nonumber
\end{equation}
with $\avg{\T(t)} = 0$ and $\avg{\T_i(t) \T_j(t')} = 2 \gamma k_BT \delta_{ij} \delta(t-t') {\bf I}$, leading to $\avg{ \mathbf{X}(t)} =0$.
By computing the mean squared displacement, we determine the diffusion ratio,
\begin{equation}
\frac{D_L(\tilde{q}) }{D_S} = \frac{C^2_{\mathrm{A}}(\tilde q) + C^2_{\mathrm{B}}(\tilde q)}{\left[C_{\mathrm{A}}(\tilde q) + C_{\mathrm{B}}(\tilde q)\right]^2}\, .
\label{eq:diffratio} 
\end{equation}

\begin{figure}
\centering 
\includegraphics[width=\columnwidth]{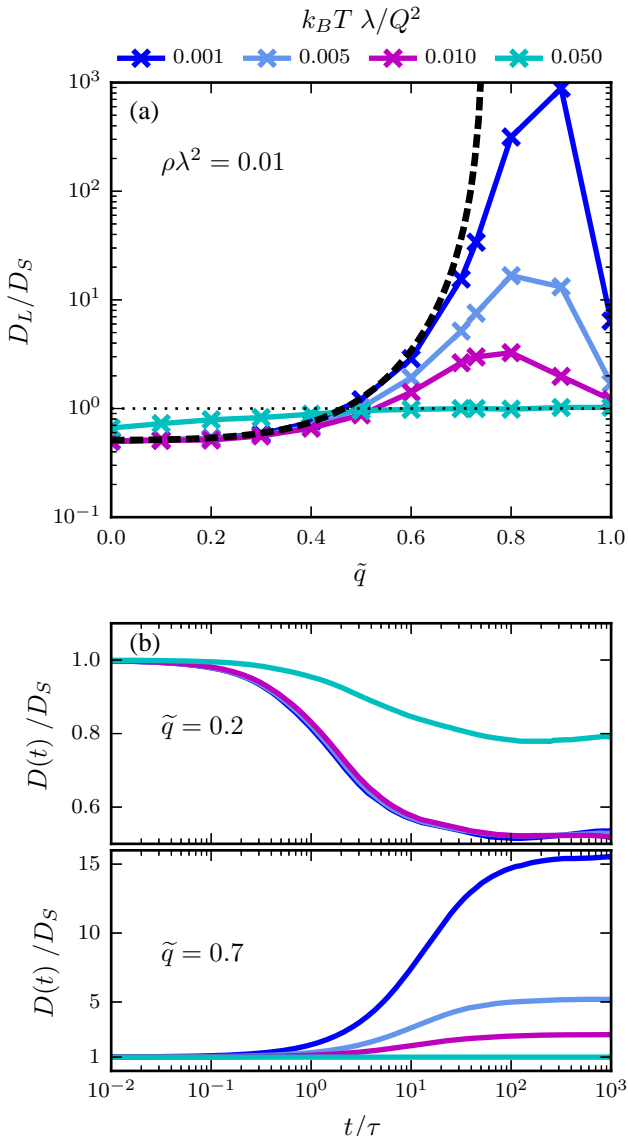}

\caption{(a) Ratio of the long-time to short-time diffusion coefficients, $D_L/D_S$, obtained from the 
time dependent diffusion coefficient $D(t)$ for given finite temperatures and number density $\rho \lambda^2 = 0.01$. 
The horizontal dotted line indicates the transition between active fluids 
($D_L/D_S>1$) and solids ($D_L/D_S<1$). The dashed line represents the analytical solution, the solid lines and symbols show numerical results.
(b) Temporal evolution of $D(t)$ normalized by $D_S$ for two chosen relative wake charges $\tilde q = 0.2$ and $\tilde q = 0.7$ with $\rho \lambda^2 = 0.01$ and varied finite temperature.
}
\label{diff_of_temp}
\end{figure}

\noindent
The result of Eq.~\eqref{eq:diffratio} is presented in Figs.~\ref{figMSD} and ~\ref{diff_of_temp} by the black dashed line.

%\begin{figure}
%\centering 
%\includegraphics[width=\columnwidth]{v_c_tau_withtemp1}
%\caption{Characteristics of active fluids for given finite temperatures: average particle velocity $\langle v \rangle$, alignment parameter $c$, and
%decay time of doublets $\tau_D$, plotted versus the relative wake charge $\tilde q$ for the number densities $\rho \lambda^2= 0.1$ and $\rho\lambda^2 = 1$. The horizontal lines in panel (c) and (f) indicate the threshold value of the decay time ($10^3\tau$).}
%\label{CharacterTemps}
%\end{figure}

\begin{figure}
\centering 
\includegraphics[width=\columnwidth]{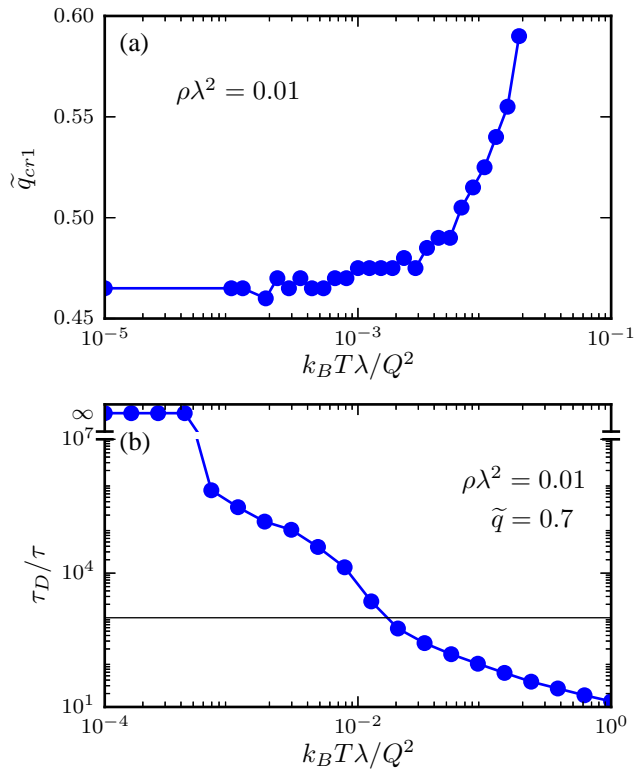}
\caption{(a) Temperature dependence of the onset of activity $\tilde q_{cr1}$ (determined from the condition  $D_L / D_S > 1$) in simulations with $\rho \lambda^2 =0.01$. 
At low temperatures it approaches the analytically derived value of $\tilde q_{cr1} \simeq 0.46$. 
(b) Temperature dependence of the doublet decay time $\tau_D$ for simulations with $\rho \lambda ^2  = 0.01$ and $\tilde q = 0.7$.
If no pair splits during the simulation time, $\tau_D$ is set to infinity.
The horizontal line shows the stability criterion of long-living active clusters. %$10^3\tau$.
}
\label{tau_and_q_of_temp}
\end{figure}

Finally, we demonstrate the impact of finite temperatures in many-body simulations.
In Fig.~\ref{figMSD}, we compare the analytical results for the diffusion coefficient, Eq.~\eqref{eq:diffratio}, 
with the numerical results obtained for number density $\rho \lambda^2= 0.01$
using the time-dependent
diffusion coefficient $D(t) =
\frac{1}{4t} \langle  |\R(t) - \R(0)|^2  \rangle$ plotted in \fig{figMSD}
saturates towards a long-time diffusion coefficient  $D_L$ at long times
which is naturally normalized by the short-time coefficient $D_S =\lim\limits_{t \rightarrow 0}{D(t)}=k_BT/\gamma$.

For intermediate times there is either a sub-diffusive regime due to particle caging, or a ballistic regime arising from the
emerging activity~\cite{Howse_2007,ZhengBtH2013}. As discussed above (see \fig{figtheory}), in dilute systems the activity
at finite temperatures is expected to set in at $\tilde q>\tilde q_{\rm cr1} (\simeq 0.46)$. From \fig{figMSD} we see that
for $\rho\lambda^2=0.01$ the transition to active fluids, $D_L/D_S >1$, indeed occurs near this value. The long-time
diffusion increases over several orders of magnitude as a function of nonreciprocity $\tilde q$. Even in dense colloidal
fluids (at $\rho\lambda^2=1.75$) the ratio $D_L/D_S$ exceeds 5, implying that there is an enormous diffusivity relative to
the case of infinite dilution. As revealed by the snapshots in \fig{figv}, this is mainly due to significant local alignment
in the fluid, which allows for an efficient traveling of active doublets.

One can see that the onset of activity, defined by the diffusion ratio $D_L/D_S>1$, remains practically unchanged at all temperatures, whereas the asymptotic deviation of $D(t)/D_S$ from unity decreases with $T$, see Fig.~\ref{diff_of_temp}.
%The effect of the temperature on the characteristics of an active fluid is shown in Fig.~\ref{CharacterTemps}.
%While the observed velocity increase is due to a trivial increase in the thermal motion,
%one can naturally see that at higher temperatures the stability is gradually decreasing.
Naturally, at higher temperatures the stability is gradually decreasing. 
The explicit dependence of the activity onset $(\tilde q_{cr1})$ and pair stability $(\tau_D)$ on the temperature is shown in Fig.~\ref{tau_and_q_of_temp}.

\section{Conclusion}
\label{sec:conc}

In conclusion, we have shown that in two-dimensional systems with wake-mediated interactions a rich variety of
self-organization phenomena occur. In the zero-temperature limit, The nonreciprocal forces exerted by wakes
generate a complex diagram of steady states. In particular, we showed the formation of active units -- bound particle pairs,
having interesting similarities with permanently active Brownian particles -- and the realization of unusual melting
scenarios. At finite temperatures we identified regimes of anomalously high diffusion. The ability of particles with the
wake-mediated interactions to form active units, the unusual melting  and the unique diffusive behavior make such systems
interesting for many fields of research. We encourage scientists in the field of colloidal dispersions or complex plasmas to
realize the experiments where these theoretical predictions can be verified.

\acknowledgments
The authors acknowledge support from the European Research Council, under the European Union's Seventh
Framework Programme, ERC Grant Agreement No. 267499, and from the Russian Scientific Foundation, Project No. 14-43-00053.
A.K. gratefully acknowledges financial support through a Postdoctoral Research Fellowship (KA 4255/1-1) from the Deutsche
Forschungsgemeinschaft (DFG).

\bibliography{ref}

\end{document}